\documentclass{article}

{}
{}
{}

\usepackage{multirow}
\usepackage{color}
\usepackage{amsmath} 
\usepackage{amsthm} 
\usepackage{cite} 
\usepackage{hyperref} 
\usepackage{graphicx}
\usepackage{algorithmic} 
\usepackage{hyperref}
\usepackage{authblk}
\usepackage[margin=1.2cm]{geometry}
\usepackage{multicol}
\usepackage{blindtext}
\usepackage{epstopdf}
\newcommand{\review}[1]{{\textcolor{black}{#1}}}

\begin{document}

\title{\textbf{An Impact and Risk Assessment Framework for National Electronic Identity (eID) Systems}}

\author[1]{Jide Edu}
\author[1]{Mark Hooper}
\author[1]{Carsten Maple}
\author[1]{Jon Crowcroft}

\affil[1]{{The Alan Turing Institute}}

\date{}

\setcounter{Maxaffil}{0}
\renewcommand\Affilfont{\itshape\small}

\maketitle

\begin{abstract}
Electronic identification (eID) systems allow citizens to assert and authenticate their identities for various purposes, such as accessing government services or conducting financial transactions. These systems improve user access to rights, services, and the formal economy. 
As eID systems become an essential facet of national development, any failure, compromise, or misuse can be costly and damaging to the government, users, and society. Therefore, an effective risk assessment is vital for identifying emerging risks to the system and assessing their impact.
However, developing a comprehensive risk assessment for these systems must extend far beyond focusing on technical security and privacy impacts and must be conducted with a contextual understanding of stakeholders and the communities these systems serve.
In this study, we posit that current risk assessments do not address risk factors for all key stakeholders and explore how potential compromise could impact them each in turn. In the examination of the broader impact of risks and the potentially significant consequences for stakeholders, we propose a framework that considers a wide range of factors, including the social, economic, and political contexts in which these systems were implemented. This provides a holistic platform for a better assessment of risk to the eID system.

\end{abstract}

\begin{multicols}{2}

\section{Introduction}
In this age of digital interaction, the ability to prove identity digitally has become increasingly crucial and valuable. 
Many governments around the world are now using electronic identification (eID) systems to assert and prove residents' identities to facilitate the delivery of e-services, welfare and benefits. Such systems have become important development initiatives that enable sustainable development and have helped empower citizens by enhancing their access to rights, services, and formal economy \cite{Joseph_2018}. These systems make it easier to provide a range of services that broaden financial inclusion, unlock economic value, increase access to social safety nets, and raise gender equality by offering a secure and precise method for identifying the population \cite{FATF_2020}. Furthermore, with eID systems, nations can comprehensively understand their citizens' political, educational, and economic behaviour to better plan and address their needs \cite{RODRIGUEZ201564}.

As eIDs are becoming an important facet of national development, any failure, compromise, or misuse of the system could be costly and damaging to the government, users, and society. There have already been reports of security and privacy incidents in media involving eID systems, such as in the case of Aadhaar, where over 200 government websites exposed Indian citizen data to the public \cite{Tiwari_and_Agarwal_2022}.
More importantly, 
these systems hold massive data belonging to millions or even billions of individuals (such as India's Aadhaar system \cite{Aadhaar_system}), making them interesting targets for adversaries. Hackers, cybercriminals, insider threats, and nation-states all pose enormous threats to the eID system \cite{ID4D_threats}.

Furthermore, existing studies have demonstrated a set of unique challenges associated with these systems owing to their scale \cite{Tiwari_and_Agarwal_2022}. For instance, extensive data logging means that they can be used to profile registered users easily. Researchers have proposed various techniques for protecting data at scale, for example, using differential privacy techniques on the anonymised log, but it has become apparent that this is not a one-size-fits-all solution \cite{264108}. Thus, compromising such a system could provide attackers with access to valuable and sensitive information, necessitating the need to develop a robust understanding of potential risks.

An excellent starting point for securing a sociotechnical system, such as an eID system, is focusing on its risks \cite{renn_2008}.
However, focusing on risk requires considering the impact of potential threats instead of simply identifying all possible threats \cite{HORLICKJONES199879}.
Despite the fast-growing research on identification system security and privacy issues, there is a lack of research on how to systematically assess the impact of successful compromise on eID systems. This study aims to fill this research gap by answering the following main research question: How can we systematically assess the impact of compromise on eID systems?
There are many eID systems, all of which are contextual in their design and operation \cite{marsman_2022}. These systems are characterised by path dependency \cite{Kubicek_2010}, their context \cite{Brugger2015RaisingAO}, and the culture they are part of \cite{Pappas_2019}. 
Like most complex socio-technical systems, 
the impacts of risk vary across the different stakeholders \cite{Mir_2022}. Any failure of critical system goals affects stakeholders differently based on what they value.

In this paper, we present a framework for assessing the risks and impacts of a compromise on eID systems. We identify the key stakeholders in eID and discuss their values (Section \ref{sec:stakeholders}).
Based on the common decomposition of risk into impact and likelihood, we discuss possible risk impact areas for assessing the impact on the stakeholders (Section \ref{sec:impactQuant}). We further discuss in Section \ref{sec:overviewframework} how these impact areas can be used to conduct an effective risk assessment of eID systems.
%
We focus on national eID systems in particular: those designed to serve a wide range of purposes, including but not limited to population registers, unique identification numbers, and national identification cards. These systems typically aim to pursue public policy objectives such as streamlining public administration, increasing security, managing services, or public governance. Examples include the Estonia identity system \cite{oecd2019digital}, which offers eID to allow its citizens to vote, submit tax claims, and check medical records, and the Aadhaar system \cite{Aadhaar_system} that provides unique identification to Indian citizens and lets them access government services.

This paper makes the following key contributions:
\begin{itemize}
    \item We identify key eID stakeholders and their values that need to be protected against risk.
    
    \item We propose a multi-stakeholder impact assessment approach, allowing independent assessment of varied risk impacts on eID stakeholders.
    \item We implement a prioritisation technique to account for the contexts of use of the eID system and the differences in what stakeholders value. 
\end{itemize}

\section{Challenges}
\label{sec:method}
There are numerous challenges in evaluating the impact of compromise on identification systems. First, there is the problem of scarce data. 
Only a few governments or development partners have rigorously assessed the effects of compromise on national identity systems. 
Even when they do,
the assumptions and figures behind the estimations are usually not publicly available \cite{Mahasen717}. Besides, governments and identity solutions providers may have incentives to downgrade the impacts \cite{guardian_2015}.

In addition, it is challenging to define uniform risk impact evaluation metrics across various sectors and agencies because the design and use of identification systems are multilayered and rarely comparable across nations \cite{worldbank_public}. 
%
While cyber risk impact has usually been assessed from the three central technical aspects of information security: confidentiality, integrity and availability, the specific attributes of eID systems, as opposed to IT systems, often prevent the straightforward application of these risk impact assessment criteria. Moreover, focusing only on technical aspects is insufficient to justify investment in security \cite{OWASP_risk_rating} and provides no insight into stakeholder impacts.

Likewise, measuring risks to complex systems like identification systems is complicated. The far-reaching and diffuse nature of the effects of compromise presents greater issues for assessment. For instance, in relying private parties such as financial institutions, identity touches nearly every transaction that involves an exchange of value or trusts \cite{FATF_2020}, and potential impact could be identified for almost every corner of the sector.
Thus, we are limited in our ability to create a valid quantitative impact evaluation model for the identification system. 
Instead, this paper uses existing literature on digital identity, news articles and reports to draw some initial conclusions and proposes an analytical framework for assessing the impact of compromise on an eID system.

\section{Key eID Stakeholders}
\label{sec:stakeholders}
In the context of eID, stakeholders are individuals and organisations involved in creating, maintaining, and using an ID system throughout the identity lifecycle. To better capture the impact of a compromise on stakeholders, we need to first understand who these stakeholders are and their roles. 

The eID scheme comprises three primary stakeholders \cite{Joseph_2018}. The provider of the eID scheme (the government), the individuals being identified, and the organisations relying on it for customer identification.

\noindent\textbf{Government}: government agencies are the primary providers of the ID systems. These include population registrars, National ID agencies, civil registrars, etc., that register people in the ID system and issue and manage credentials. In addition, they are responsible for managing and updating the identity data, resolving disputes and offering authentication and verification services at various levels of assurance \cite{ID_Stakeholders}. Other government agencies also use these ID systems to interact with people or provide functional ID systems.

\noindent\textbf{Individuals}: These are the people at the centre of the Id systems. They are the system subjects whose personal data are collected. They are entitled to exercise appropriate control over their data collection, storage, and dissemination. Likewise, they are the people who use their credentials and proof of ID to access rights and services \cite{ID_Stakeholders}. Therefore, building an ID system capable of advancing development objectives must begin with an understanding of and response to people's ID-related needs and concerns~\cite{Brugger2015RaisingAO}. 

\noindent\textbf{Relying private organisations}:
Many organisations use government ID systems to identify their customers, such as requiring government-issued credentials to open bank accounts, register SIM cards, or set up credit reporting systems. This stakeholder uses ID providers' platforms, credentials, and services to authenticate or verify the end-users identities. 

Other stakeholders such as civil society, international organisations and development partners~\cite{ID_Stakeholders} also help people use eID systems, advocate for inclusion, and serve as sources of critical feedback for eID system planning and implementation. 

\section{Impact Assessment Factors}
\label{sec:impactQuant}
Assessing the impact of risk on different eID stakeholders involves considering the potential consequences of the risk event for all stakeholders affected by it.
As a first step, we need to establish the key factors to consider when evaluating the impact of risk on the identity system's mission and objectives. This entails establishing a set of evaluation criteria against which the consequences of a realised risk can be assessed \cite{OctaveAllegro_2007}. Besides, defining evaluation criteria is essential to security risk management as security threats must have a clearly defined impact before they can be considered system risks \cite{Peltier_2001}.
%
However, because the consequence of a single risk event to these systems can have multiple potential impacts on various stakeholders depending on their value, it is paramount to establish risk impact areas that align with these values.

We distinguish three types of impact on an eID system: ``impact on the government'', ``impact on the end-users''; and ``impact on the relying private organisations''. 
These are further discussed below:

\subsection{Impact on the government}
To understand the government value attributed to implementing an eID system and which areas the risk might impact, we analyse the existing literature on public value.

\end{multicols}

\twocolumn

The authors in \cite{Cresswell_2006} proposed a Public Return On Investment (PROI) framework to evaluate the value of government investments in information technology. The framework identifies two public return sources: value to the public that comes from improving the government as a whole from the point of view of the citizens and the value that comes from providing specific benefits directly to individuals, groups, or the general public.
PROI defines six main government values based on the different impacts government investments in information technology can have on public stakeholders' interests. 
These are \textit{financial, political, social, ideological, stewardship, and strategic values.} Another framework to value various government initiatives proposed in \cite{Foley2006USINGTV}, Value Measuring Methodology (VMM), also identified five key government values: direct value to customers, social and public value, financial value to the government, operational and foundational value to the government, and strategic and political value.

Using these different value measurement dimensions, we distinguish seven key risk impact areas for the government, as shown in Table \ref{tb:impactGov}: \emph{reputation, economic, social, political, operational, physical, and rights}.

\begin{table}[ht]
\centering
\caption{Derive government impact areas from PROI and VMM frameworks}
\vspace{1.2mm}
\label{tb:impactGov}
\scalebox{0.9}{
\begin{tabular}{l|l|l}
\textbf{Impact Category} & \textbf{PROI} & \textbf{VMM}                      \\ \hline
Economic   & Financial & Financial   \\ \hline
& Ideological  &   \\ \cline{2-2}
\multirow{-2}{*}{Reputation}    & Stewardship    & \multirow{-2}{*}{}      \\ \hline
Social    & Social   & Social and public    \\ \hline
 & Political   &  \\ \cline{2-2}
\multirow{-2}{*}{Political}  & Strategic    & \multirow{-2}{*}{Strategic and political } \\ \hline
Citizen rights  &   & Customer direct value    \\ \cline{1-2} \cline{3-3} 
Operational    & \multirow{-2}{*}{} & Operational \\
\end{tabular}}
\end{table}

\noindent\textbf{Reputation Impact}:
The reputation impact is related to the citizen's trust in the government.  
It is essential that an eID system is widely adopted. One of the critical elements for wide adoption is public trust in the system \cite{Halperin_2012}. Trust in the government is essential for the public acceptance of eID systems. A lack of trust will prevent citizens' participation, jeopardising the identity program, thereby widening rather than closing the humanitarian divide.
Apart from harming users' trust, there could also be damage to trust relationships with other governments or non-governmental entities.
Thereby undermining the eID system's usefulness by making it difficult for organisations to rely on them for secure authentication and authorisation.
Thus, measuring the impact of any compromise on trust in the government and the ID system is critical.

\noindent\textbf{Economic Impact}:
This includes all direct and indirect financial damages experienced by the ID agency. The financial implications can be assigned according to the ID agency’s financial loss in terms of the \emph{one-time financial cost} or \emph{operating cost} of investigating and rectifying the compromise. Likewise, we could measure the economic impact in terms of \emph{revenue loss}, as governments use the system to improve their revenue through better tax collection \cite{SULE2021101734, Joseph_2018}. 
In addition, the eID system offers government financial gains by limiting potential leaks in government benefit programs and eliminating bloated civil service wages from ghost workers \cite{Joseph_2018}. A compromise could result in financial loss to the government. 
For example, in 2011, claims filed in the US under compromised identities resulted in fraudulent unemployment benefits payments totalling \$3.3 billion \cite{huffpost_2013}. 
The economic impact could also be in terms of economic growth and innovation, budget impact, workforce loss, diminished foreign investments, or fines and legal penalties \cite{10.1093/cybsec/tyy006}.

\noindent\textbf{Rights}:
eID systems allow governments to perform their obligations under international human rights law by giving every citizen the right to be recognised as a person and to be treated equally before the law \cite{worldbank_public}. 
The government leveraged identification systems to create transparency, fairness, and well governed services.
These systems are carefully designed for inclusion, fairness, equity, and accessibility. This is because the government aims to ensure equal access to digital services for all citizens. Hence, these systems are implemented to facilitate access to public and private services, particularly for less-privileged parts of the population \cite{worldbank_public}.
The size and quality of readily available services in a broader eGovernment ecosystem are some of the literature-identified adoptions and acceptance drivers \cite{SAMUEL2020408}.
Impact evaluations can attempt to determine the number of people/end users who have been excluded and do not have access to the identity service. The implementation of robust and inclusive identification systems also correlates with broader levels of effective governance.

\noindent\textbf{Social Impact}:
One of the fundamental principles of identification for sustainable development is to create an interoperable platform that is responsive to the needs of various groups of users \cite{Lother_2021}.
In addition to fairness, equity, and inclusiveness, measuring the effectiveness and efficiency of service delivery that utilises an identification system is also important.
Like the private sector, the government can be considered a service provider for its customers--- citizens. Its ultimate objectives are to improve overall social welfare and satisfy citizens' demands and needs \cite{RAUS2010122}.
A compromise could cause damage to or hinder a critical infrastructure sector impacting the way people live and interact in society.

\noindent\textbf{Political Impact}:
A compromise could impact public influence and disrupt political processes, threatening national security. 
Countries have used eID to prevent vote rigging. For example, Nigeria uses eID to authenticate voters using biometrics, preventing approximately 4 million duplicate voters \cite{Joseph_2018}.
In this instance, a compromised eID system could make it more likely that election results will be disputed, increasing the risk of election violence and the associated human and financial costs.

\noindent\textbf{Operational Impact}:
Operational impact refers to the operational damage that affects the mission capability of the government and its effectiveness.
Suppose the compromise eID is used for a non-time-sensitive transaction, the impact of a limited-duration availability compromise on the government's objective and public confidence will be minimal in most cases. However, the availability of time-sensitive information is less likely to be restored before significant harm is done to the agency or public welfare. The impact of compromise on government operations can also be evaluated based on the number of teams that deal with risk events.

\noindent\textbf{Physical Impact}:
Physical impact refers to the damage or destruction to the government’s physical properties, assets or resources. This can include staff, equipment, buildings, and other infrastructure used to support an eID project.

\subsection{Impact on relying private organisations}
Private organisations rely on digital identity services for identification purposes. However, unlike in public organisations, the value attributed to private parties differs owing to diverse needs and requirements.
In a private market, the ultimate goal of a business is profit and shareholder value maximisation \cite{RAUS2010122}.
Most private sector valuation methods are inextricably linked to economic value and are measured in monetary terms. 
However, besides financial value, there are other values, such as the operational value of increasing productivity, service quality and compliance, and the social value of improving customer satisfaction \cite{RAUS2010122}.
Next, we discuss how a compromise in eID systems might affect these values.

\noindent\textbf{Economic Impact}:
Relying private organisations benefit from a robust and inclusive identification system. 
The ability to accurately identify customers, mitigate fraud risk, and reach new markets for trusted users and partners is essential to private organisations. 
A compromise could result in a regulatory fine, compensation payments, disrupted turnover, PR response costs, and reduced profits. 
In some cases, relying private parties may also face legal liabilities if they are found to have failed to adequately protect personal data or to have used inadequate security measures when relying on an eID system.

\noindent\textbf{Operational Impact}:
Operational impact here refers to the extent of the breach on the business mission and objectives of the private organisations or the effect of the service interruption. 
Besides, private organisations that rely on eID systems are service providers for the citizens \cite{RAUS2010122}.

\noindent\textbf{Physical Impact}:
The physical impact can be assessed based on damage or destruction to the organisation’s physical property or resources. This includes staff, equipment, buildings, and other types of infrastructure.

\noindent\textbf{Reputation Impact}:
A compromise can also impact the reputation of relying private parties. 
The reputation impact is related to the citizen's trust in the organisation. 
For example, if a breach occurs through relying private parties, they may experience public backlash and loss of customer trust.
Reputation impact could be measured in terms of damage to public perception or the level of media scrutiny.
How an organisation handles a compromise could also impact its reputation.

\noindent\textbf{Social Impact}:
The social impact here refers to the effect of compromise on the people and communities in which the relying party operates and serves. This area could be evaluated by the impact on the organisation's activities, such as the jobs it creates and the products or services it provides.
Besides, many organisations are increasingly focused on maximising their positive social impact by implementing sustainable business practices, supporting local communities, and engaging in corporate social responsibility initiatives.

\subsection{Impact on End-Users}
End-Users are more concerned with the identity system's transparency, usability, accessibility, availability, privacy, and security \cite{Brugger2015RaisingAO}. We distinguish six key risk impact areas for end-users: rights, physical, privacy, social, psychological, and economical. These are further discussed below.

\noindent\textbf{Privacy Impact}:
One of the most valuable assets of an eID system is personal data. Therefore, compromising such a system will impact users' privacy, even though this may affect users differently depending on what they value. 
An important contributing factor is what has happened to the data: has it been made public, changed, or used to make decisions that affect people? If exposed, to whom and what harm could and would they cause? This will help to better measure the severity of violations.
As end-users are people whose personal data are collected, they are the ones that will be directly affected by privacy violations. The fact that only individuals (end-users) can directly experience a privacy problem is especially challenging for assessing the impact.

\noindent\textbf{Right}:
Not only does eID enable people to exercise their rights in an inclusive manner, it also holds them accountable for their obligations.
For instance, eID can make it easier for citizens to access government services, especially those excluded because of the difficulty in obtaining physical documents or living in remote areas \cite{Joseph_2018}.
The impact of an identity breach could be evaluated by what service restriction it can cause, including those we consider fundamental human rights. Is it a breach that causes the risk of right violations that would not ordinarily be subject to enforcement efforts or those that are particularly important to enforcement programs?

\noindent\textbf{Social Impact}:
This impact area measures how a breach affects society through the options available to end users.
A compromised identity system has significant repercussions for citizens, who may be denied access to numerous services, and society at large, which is unable to develop or organise effectively \cite{Lother_2021}.
In an eID system, the capability and opportunity should be equal, and the process of choosing and seeking a specific opportunity should also be fair to all end-users \cite{marsman_2022}. A compromised national identity system could impact end-users by introducing barriers to access and usage, thereby creating disparities in the availability of information and technology.

\noindent\textbf{Psychological impact}:
There could be psychological impacts on end-users, including, but not limited to, suffering, inconvenience, and distress.
This might occur due to not having access to a critical service, especially if this happens over a long period or on more than one occasion. It could also be the time the end-users spent and their efforts to seek a resolution for the breach. End-users might also suffer from anxiety, upset or stress due to compromise. For example, criminals may commit non-financial crimes in another person's name (identity theft), for which the victim is held accountable and suffers the consequences \cite{Harrell_2016}. Sometimes, the psychological impact lasts much longer than the financial impact.

\noindent\textbf{Physical Impact}:
It is crucial to assess the impact the compromise might have on people's lives, particularly if the breach could harm them.
A compromise resulting in identity theft that allows criminals to falsify registration details could support child labour and human trafficking \cite{doi:10.1080/02681102.2020.1840325}, which has significant implications for individuals' safety. Moreover, exclusion from life essential programs such as food security schemes can lead to starvation and death \cite{10.1145/3476056}. People can also be exposed to physical harm, as was the case during the Rwandan genocide when roadblocks used ethnicity written prominently on Rwandan ID cards to determine who would be murdered \cite{Fussell_2001}.

\noindent\textbf{Economic Impact}:
Economic impact refers to the financial loss suffered by users in terms of the cost of services or fraud instead of loss of profit. For example, end-users can be financially impacted if an attacker uses their identity for unauthorised financial transactions \cite{FATF_2020}.

\section{Overview of the framework}
\label{sec:overviewframework}
\begin{figure}[!ht]
    \centering
        \vspace{-1mm}
    \includegraphics[width=\columnwidth, trim=0 0 40 0, clip]{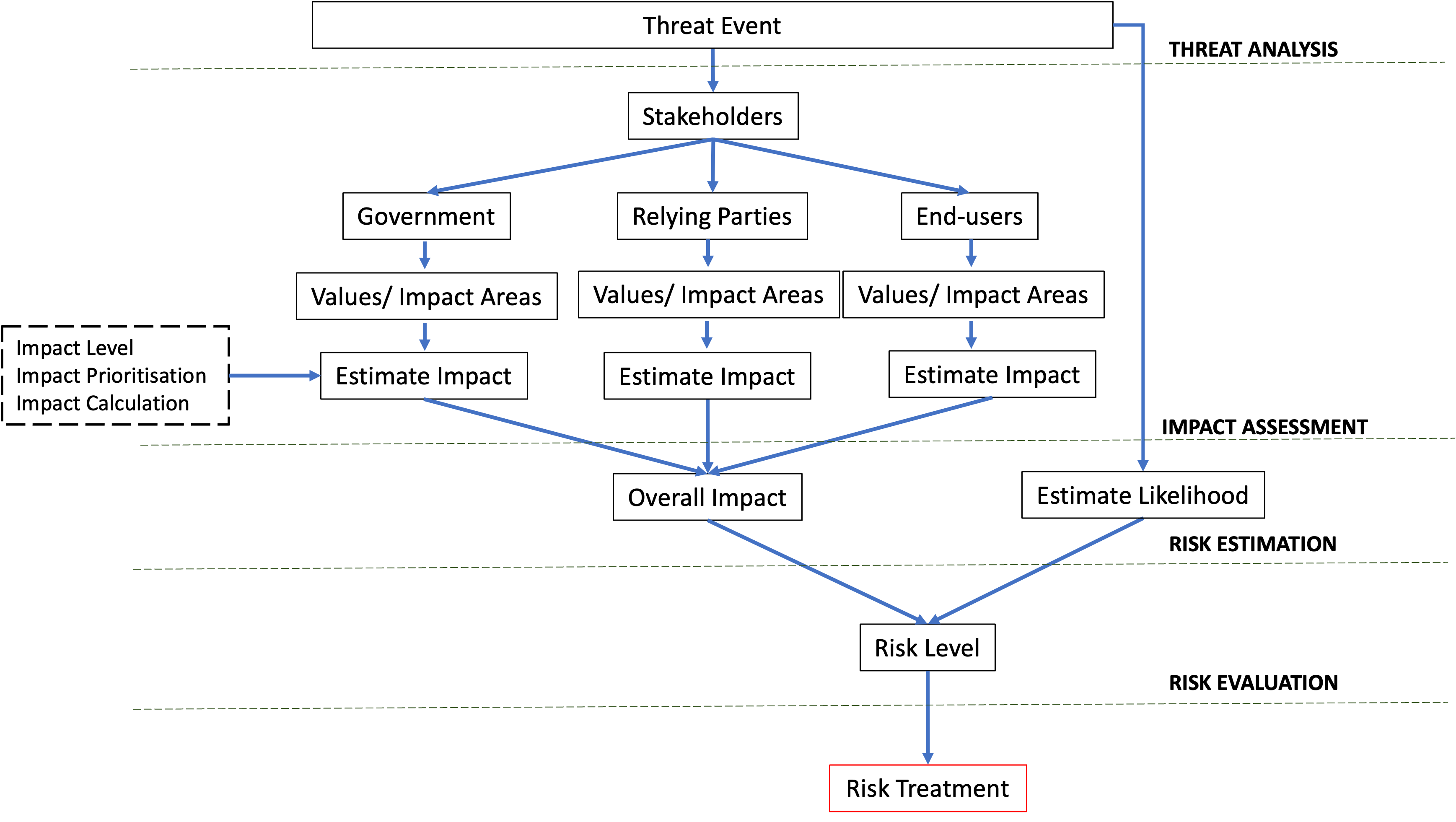}
    \vspace{-5mm}
    \caption{Workflow of the risk assessment framework}
    \label{fig:overview}
     \vspace{-5mm}
\end{figure}

Risk assessment is an essential part of best risk management practice.
%
The core idea behind risk assessment is to employ analytical and structured methods to gather data, opinions, and evidence about what is at risk, the likelihood of unwanted events, and a measure of the impacts. 
There are many different approaches to risk assessment (we refer readers to \cite{ENISA_risk} for detailed analysis). Our framework is based on these standard methodologies and aims to compute the risk for the eID system considering its specific attributes. As depicted in Figure \ref{fig:overview}, the risk assessment framework consists of the following phases, i) context establishment, ii) risk identification, iii) risk estimation, and iv) risk evaluation, which we will next discuss in detail.

\subsection{Context Establishment}
In this phase, the scope and broad context of risk assessment are defined.
By establishing a context for risk assessment, we can understand the external and internal factors that could affect the eID system's ability to achieve its mission, goals, and objectives. 
During this phase, the granularity levels of the eID system to be evaluated are defined.
Basic risk measurement criteria are established as part of the context establishment process. Setting the basic criteria involves defining stakeholders' values, risk impact measurement, risk evaluation, and risk acceptance criteria.
This phase sets the stage for the risk identification phase since defining the eID system objectives is a prerequisite for identifying risks.

\subsection{Risk Identification}
Risk identification in digital identity systems involves identifying and documenting the potential risks that may impact the identity system. This process helps to establish a clear understanding of which threats target which eID assets and what might happen if those attacks succeed \cite{NIST_2012, OctaveAllegro_2007}.
Several methods can be used to identify potential risks to eID systems. One popular technique is brainstorming, in which individuals with knowledge and expertise in digital identity systems generate a list of potential risks through open discussion and idea sharing \cite{ISO_31000}.
Root cause analysis is another technique that involves identifying the underlying causes of potential eID risks and eliminating or mitigating these causes. This can be performed using tools such as 5-Whys \cite{murugaiah2010scrap} or fishbone diagrams \cite{LUO2018296}. 
Risks can also be identified through a scenario analysis by considering future events and the possible risks that may result from them. This can be conducted using tools like decision trees or Monte Carlo simulations \cite{ISO_31000}. 

\subsection{Risk Estimation}
Risk estimation in an eID system is the process of determining the impact and likelihood of a particular eID risk \cite{renn_2008}. This consists of two fundamental processes: assessing risk impact and determining the likelihood of risk.

\subsubsection{Estimate the risk impact}
This process determines risk consequences. It involves three steps: identifying impact levels, prioritising impact areas, and calculating impact scores.

\noindent \textbf{Impact level identification:}
Without a reference standard for comparison, comparing and aggregating risks across an eID system is impossible.
Many frameworks \cite{NIST_2012,OctaveAllegro_2007} rate impact on Likert scales, e.g. from ``very low'' to ``very high''. For example, the NIST guidelines \cite{NIST_2012} provide examples of adverse impacts, such as financial loss and harm to operations, assets, or individuals, and explain how the Likert scale should be used to determine the expected extent of each impact. 
The scales comprise rating levels and definitions that foster consistent interpretation and application by different users.
We represent the impact areas identified in Section \ref{sec:impactQuant} by 
a three-point scale labelled ``significant'', ``moderate'', and ``minor''.
A numeric impact value range can also be added to the scale point as follows: ``Significant'' – (70-100), ``Moderate'' – (31-69), and ``Minor'' – (0-30) for semi-quantitative analysis. These relative values may be amplified depending on the required granularity to visualise the risk metrics.

\noindent \textbf{Risk impact areas prioritisation:}
In addition to evaluating the extent of an impact, there is a need to prioritise the impact areas from most important to least important based on their importance to the identity system operation and objectives. For example, an identity system developed for voter registration might prioritise citizens' rights over other impact areas. Risks that impact the citizen's rights will generate higher scores than risks with equivalent impacts and probabilities in another area.
A quantitative weight can be assigned to each impact area where the most important is given the highest value, and the least is assigned the lowest value \cite{OctaveAllegro_2007}.
\review{As part of context establishment, stakeholder requirements analysis can help in defining the ranking process and how the weights should be assigned.}

\noindent \textbf{Impact score calculation:}
The anticipated loss for stakeholders is estimated by the impact level of a particular risk.
We define the impact of risk on any given eID system by the extent of the risk (impact value) and the impact area priority rank (quantitative weight) set by the user. 
\review{This definition considers the contexts of use of the eID system and the differences in what stakeholders value.}
It can be mathematically expressed as follows:

\begin{equation}
    \bar{i} = \sum_{n=1}^n{w_n\cdot C_n}
\end{equation}

\noindent where $C$ represents the impact value and $w$ represents the quantitative weight of the impact area. For instance, expressing this equation using the impact areas identified in Section \ref{sec:impactQuant}, the risk impact on relying private parties $i_r$ can be represented with the following simple linear equation.

\begin{equation}
i_r = eo_rw_1 + op_rw_2 + ph_rw_3 + re_rw_4 + so_rw_5
\end{equation}

That is, using the impact area identified for the relying private parties, the risk impact score $I_r$ is computed using the allocated impact value and weight for economic (eo), operational (op), physical (ph), reputation (re) and social (so) impact areas. $w_1$ to $w_5$ in the equation represents the quantitative weight value of the impact areas set by the ``user''. 

The overall impact score $I_{sum}$ for the risk on stakeholders can then be computed by aggregating the impact on government, relying private parties and end-users as follows.



\begin{equation}
    \label{eqn:overallimpact}
    I_{sum} = i_g + i_r + i_e
\end{equation}

Where $i$ is the computed impact value, the indices g, r, and e stand for the government, relying parties and end-users, respectively. 

\subsubsection{Estimate the likelihood}
In this phase, the likelihood of the risk occurrence is assessed. 
Estimating the likelihood involves analysing the probability that a risk event will occur based on factors such as:
\begin{enumerate}
 \item Vulnerability: How vulnerable the eID system is to the feared event based on its configuration, software, etc.
    \item Threat actor capability: The expertise, knowledge about the target eID system, tools, and resources an attacker may have to exploit the vulnerability \cite{NIST_2012}.
    \item Motivation: The motive behind the attack, such as financial gain, political or ideological reasons, or revenge.
    \item Historical data: Historical data that can provide insights into the likelihood of specific types of attacks \cite{ISO_31000}.
\end{enumerate}
\noindent We can rate the likelihood as ``high'', ``medium'', or ``low''. A ``high'' likelihood rating exists if the threat source is sufficiently capable and highly motivated, and the measures to guard against the vulnerability from being exploited are ineffective. If the threat source is competent and motivated and there are measures in place that could prevent the vulnerability from being exploited successfully, this is a ``medium'' rating. There is a ``low'' rating if sufficient controls are in place to prevent or at least hinder the exposure from being exploited or if the threat source lacks the
capability to do so.

\subsection{Risk Evaluation}
Once risks have been identified and estimated, evaluating them in terms of their potential overall impact and likelihood is important.
The combination of the likelihood and impact levels can be used to determine the risk level. 
The risk level allows us to recognise the risks that have the most significant impact on the eID system.
The risk level can be rated as significant if there is a serious and urgent threat to eID systems (risk reduction remediation should be instantaneous), elevated if there is a real threat to eID systems (risk reduction remediation should be completed within a reasonable period), and low if threats are common and generally acceptable, but may still have an impact on eID systems. Additional security measures could offer greater protection against present or future unforeseeable threats.
\review{Table \ref{tb:rrc} depicts how each risk can be rated. For instance, if the risk rate score is above 50, the risk is of grave concern to the eID system. How the risk will be rated is usually based on the stakeholders' risk appetite.
A high potential impact risk is often a great concern to decision makers, even if the likelihood is very low. Furthermore, frequent but low-impact risks can have long-term or cumulative consequences\cite{ISO_31000}.
These types of risk require consideration, as appropriate risk treatments can differ. Nevertheless, identifying and implementing appropriate countermeasures is outside the scope of the current framework.}

\begin{table}[!ht]
\centering
\caption{Risk rating calculation}
\label{tb:rrc}
\scalebox{0.86}{
\begin{tabular}{p{1.39cm}|p{0.85cm} c p{1.2cm}|p{0.53cm} c p{0.85cm}|p{2.5cm}} 
\multicolumn{2}{c}{\textbf{Impact} ($I$)} &  & \multicolumn{2}{c}{\textbf{Likelihood}~($L$)} &  & \multicolumn{2}{c}{\textbf{Risk} ($I \times L$)}                   \\ 
\cline{1-2}\cline{4-5}\cline{7-8}
Level       & Score              &  & Level    & Score                    &  & Value Range & Description                             \\ 
\cline{1-2}\cline{4-5}\cline{7-8}
Significant & 70-100             &  & High     & 1                        &  & 51-100      & Risk is of grave concern (significant)  \\ 
\cline{1-2}\cline{4-5}\cline{7-8}
Moderate    & 31-69              &  & Moderate & 0.5                      &  & 21-50       & Risk is of moderate concern (elevated)  \\ 
\cline{1-2}\cline{4-5}\cline{7-8}
Minor         & 0-30               &  & Low      & 0.1                      &  & 0-20        & Risk is of low or no concern (low)      \\
\end{tabular}}
\vspace{-4mm}
\end{table}

\subsection{Use cases}
\label{sec:usecase}

To illustrate how risk impact assessment can be conducted using the aforementioned guidelines, we explore two examples. 
\begin{enumerate}
\item Example 1: e-ID systems suffered a denial of service attack and went down for 2 minutes affecting 1\% of the population. This resulted in negative social media, and about ten thousand people could not access the government services.
\item Example 2: an identity provider server configuration error results in a data breach where sensitive users' data are exposed to unauthorised parties. It violates users' privacy, causes negative social media, and causes financial loss to people and the government. 
\end{enumerate}

We calculated the impact of the risks and derived their risk values using the formula described above.

\noindent \textbf{Impact level identification:}: for the first example, the consequence indicates direct effects on the residents psychologically and socially. As illustrated in Table \ref{tb:example1}, the risk has little or no impact on the government and relying private parties, so a value of ``minor'' has been assigned to their impact areas. For example 2, there is a significant impact on the right, reputation, economy, society, and privacy (cf. Table \ref{tb:example2}).

\noindent \textbf{Impact risk area prioritisation:} considering our use cases, the impact areas have been prioritised as shown in Table \ref{tb:example1}. The stakeholders (except those relying on private parties) considered citizens' rights to be the most crucial impact area, and physical harm was the least important. These impact areas were assigned numerical weighted values between 1 and 7 for the government, values between 1 and 6 for the end-users, and values between 1 and 5 for the relying private parties.

\noindent \textbf{Impact score calculation:} we compute the score for each impact area by multiplying the impact weight by the impact value assigned using the impact scales. 
We then compute the average impact scores $I_g$, $I_r$, and $I_e$, which represent the impact of risk on the government, relying on private parties, and end-users, respectively.
For the first example, as shown in Table \ref{tb:example1}, we obtained the impact scores of \textit{19} for the government, \textit{32} for end users and 15 for the relying parties.
Similarly, for Example 2 in Table \ref{tb:example2}, we have an impact score of \textit{75} for the government, \textit{79} for end-users, and 43 for the relying parties.

The overall impact score of the risk $ I_{sum}$ for the eID system can be computed by aggregating the risk impact scores from the government, relying parties and end-users. 
Using Equation \ref{eqn:overallimpact}, for Example 1 we have: $$ I_{sum} = (19 + 33 + 15)/3 = 22$$
For Example 2 we have: $$ I_{sum} = (75+ 79 + 43)/3 = 65$$

\noindent \textbf{Determing the likelihood:} for our examples, we assume that the likelihood of the risks occurring is high and that the threat source is sufficiently capable and highly motivated. Thus, a likelihood value of one was assigned to the risks.

\noindent \textbf{Risk evaluation:} 
%
As shown in Table \ref{tb:rrce}, the relative impact score can be used with the risk likelihood to compute the risk rating, which can be used to prioritise the identified risks base on the stakeholders' risk appetite.
For the DoS attack, we obtained a risk score of 22, representing an elevated risk level compared to the significant risk level of 65 obtained for the data breach attack that exposed the users' data.

\begin{table}[!ht]
\vspace{-2mm}
\caption{Impact measurement for example 1 (denial of service)}
\label{tb:example1}
\scalebox{0.8}{
\centering
\begin{tabular}{p{1.95cm}|p{3.3cm}|p{1.3cm}|p{0.8cm}|p{0.99cm}|p{0.75cm}}
\textbf{Government} & \textbf{Description} & \textbf{Level} & \textbf{Value} & \textbf{Weight} & \textbf{Score} \\ 
\hline
Right & Minor impact on peoples right & Minor & 25 & 7 & 175 \\ 
\hline
Reputation & Limited impact on the ID agency & Minor & 30 & 6 & 180 \\ 
\hline
Political & Minor political impact & Minor & 8 & 5 & 40 \\ 
\hline
Economic & No economic loss. & Minor & 10 & 4 & 40 \\ 
\hline
Operational & Within their Service Level Agreement of 99.9\% uptime. & Minor & 10 & 3 & 30 \\ 
\hline
Social & A minor impact on the society & Minor & 30 & 2 & 60 \\ 
\hline
Physical & No physical harm to government assets & Minor & 8 & 1 & 8 \\ 
\hline
\multicolumn{4}{r|}{\textbf{Total}} & \multicolumn{1}{l|}{\textbf{28}} & \multicolumn{1}{l}{\textbf{533}} \\ 
\cline{4-6}
\multicolumn{5}{r|}{\textbf{Impact Score}} & \multicolumn{1}{l}{\textbf{19}} \\ 
\cline{4-6}\\

\textbf{End-users} & \textbf{Description} & \textbf{Level} & \textbf{Value} & \textbf{Weight} & \textbf{Score} \\ 
\hline
Right & Minor impact on peoples right & Minor & 25 & 6 & 150 \\ 
\hline
Privacy & No privacy violation & Minor & 1 & 5 & 5 \\ 
\hline
Psychological & Long-term inconvenience or distress & Significant & 85 & 4 & 340 \\ 
\hline
Economic & No economic loss & Minor & 10 & 3 & 30 \\ 
\hline
Social & Significant impact on the society & Significant & 80 & 2 & 160 \\ 
\hline
Physical & No physical harm to individuals & Minor & 8 & 1 & 8 \\ 
\hline
\multicolumn{4}{r|}{\textbf{Total}} & \multicolumn{1}{l|}{\textbf{21}} & \multicolumn{1}{l}{\textbf{693}} \\ 
\cline{4-6}
\multicolumn{5}{r|}{\textbf{Impact Score}} & \multicolumn{1}{l}{\textbf{33}} \\ 
\cline{4-6}\\

\textbf{Relying Parties} & \textbf{Description} & \textbf{Level} & \textbf{Value} & \textbf{Weight} & \textbf{Score} \\ 
\hline
Economic & No Economic loss & Minor & 10 & 5 & 50 \\ 
\hline
Reputation & Limited impact on reputation & Minor & 20 & 4 & 80 \\ \hline
Operational & Minor  impact on operation & Minor & 10 & 3 & 30 \\ 
\hline
Social & Minor impact on the society & Minor & 25 & 2 & 50 \\ 
\hline
Physical & Minor physical harm & Minor & 8 & 1 & 8 \\ 
\hline
\multicolumn{4}{r|}{\textbf{Total}} & \multicolumn{1}{l}{\textbf{15}} & \multicolumn{1}{l}{\textbf{218}} \\ 
\cline{4-6}
\multicolumn{5}{r|}{\textbf{Impact Score}} & \multicolumn{1}{l}{\textbf{15}} \\\cline{4-6}
\end{tabular}}
\vspace{-5mm}
\end{table}


\begin{table}[!ht]
\caption{Impact measurement for example 2 (data breach)}
\label{tb:example2}
\scalebox{0.8}{
\centering
\begin{tabular}{p{1.9cm}|p{3.1cm}|p{1.4cm}|p{0.8cm}|p{0.99cm}|p{0.5cm}}
\textbf{Government} & \textbf{Description} & \textbf{Level} & \textbf{Value} & \textbf{Weight} & \textbf{Score} \\ 
\hline
Right & Serious impact on peoples right & Significant & 95 & 7 & 665 \\ 
\hline
Reputation & National media attention. & Moderate & 75 & 6 & 450 \\ 
\hline
Political & Moderate political impact & Moderate & 60 & 5 & 300 \\ 
\hline
Economic & Significant economic impact & Significant & 85 & 4 & 340 \\ 
\hline
Operational & Moderate impact on ID agency operations. & Moderate & 60 & 3 & 180 \\ 
\hline
Social & Significant impact on the society & Significant & 80 & 2 & 160 \\ 
\hline
Physical & Minor physical harm to government assets & Minor & 8 & 1 & 8 \\ 
\hline
\multicolumn{4}{r|}{\textbf{Total}} & \multicolumn{1}{l|}{\textbf{28}} & \multicolumn{1}{l}{\textbf{2103}} \\ \cline{4-6}
\multicolumn{5}{r|}{\textbf{Impact Score}} & \multicolumn{1}{l}{\textbf{75}} \\ 
\cline{4-6}\\

\textbf{End-users} & \textbf{Description} & \textbf{Level} & \textbf{Value} & \textbf{Weight} & \textbf{Score} \\ 
\hline
Right & Serious impact on peoples right & Significant & 95 & 6 & 570 \\ 
\hline
Privacy & Release of personal information to unauthorised parties & Significant & 90 & 5 & 450 \\ 
\hline
Psychological & Serious short-term discomfort, and distress & Moderate & 58 & 4 & 232 \\ 
\hline
Economic & Significant economic impact & Significant & 82 & 3 & 246 \\ 
\hline
Social & Significant impact on the society. & Significant & 80 & 2 & 160 \\ 
\hline
Physical & Minor physical harm to individuals & Minor & 8 & 1 & 8 \\ 
\hline
\multicolumn{4}{r|}{\textbf{Total}} & \multicolumn{1}{l|}{\textbf{21}} & \multicolumn{1}{l}{\textbf{1666}} \\ 
\cline{4-6}
\multicolumn{5}{r|}{\textbf{Impact Score}} & \multicolumn{1}{l}{\textbf{79}} \\ 
\cline{4-6}\\

\textbf{Relying Parties} & \textbf{Description} & \textbf{Level} & \textbf{Value} & \textbf{Weight} & \textbf{Score} \\ 
\hline
Economic & Moderate economic impact & Moderate & 45 & 5 & 225 \\ 
\hline
Reputation & Limited impact on relying private parties' reputation. & Minor & 20 & 4 & 80 \\ 
\hline
Operational & Moderate impact on operations. & Moderate & 65 & 3 & 195 \\ 
\hline
Social & Moderate impact on the society & Moderate & 71 & 2 & 142 \\ 
\hline
Physical & Minor physical harm to assets & Minor & 8 & 1 & 8 \\ 
\hline
\multicolumn{4}{r|}{\textbf{Total}} & \multicolumn{1}{l|}{\textbf{15}} & \multicolumn{1}{l}{\textbf{650}} \\ 
\cline{4-6}
\multicolumn{5}{r|}{\textbf{Impact Score}} & \multicolumn{1}{l}{\textbf{43}} \\
\end{tabular}}
\vspace{-4mm}
\end{table}


\begin{table}
\caption{Example risk rank calculation}
\label{tb:rrce}
\scalebox{0.78}{
\begin{tabular}{p{1.37cm}|l|p{0.7cm} l l|p{0.7cm} l p{0.7cm}|l} 
 & \multicolumn{2}{c}{\textbf{Impact}} &  & \multicolumn{2}{c}{\textbf{Likelihood}} & \multicolumn{1}{l}{} & \multicolumn{2}{c}{\textbf{Risk}} \\ 
\cline{2-3}\cline{5-6}\cline{8-9}
\textbf{Example} & Level & Score ($I$) &  & Level & Score ($L$) &  & Value ($I x L$) & Level \\ \cline{1-3}\cline{5-6}\cline{8-9}
\centering
1 & Minor & 22 &  & Moderate & 1 &  & 22 & Moderate \\ \cline{1-3}\cline{5-6}\cline{8-9}
\centering
2 & Moderate & 65 &  & High & 1 &  & 65 & Significant\\
\end{tabular}}
\vspace{-3mm}
\end{table}

\section{Discussion and Limitations}
\label{sec:limit}
\noindent\textbf{Impact focus measurement}
A key benefit of an impact focused risk assessment framework is that we do not need to rely on the knowledge of threats and attacks, which is necessarily incomplete. This is because the threat landscape is dynamic and it is impossible for anyone to be sure that they have complete knowledge \cite{10.1093/cybsec/tyy006}. Moreover, in an environment such as the eID system, where the threat landscape changes rapidly and novel attack patterns will continue to emerge, understanding the potential impact of the attacks on eID assets may help lessen the related uncertainty in risk management activities.
.

\noindent\textbf{Risk management strategies}:
Risk assessments alone do not necessarily reveal the appropriate mitigation strategies. Other factors such as legislation, regulations, strategy objectives, treatment options, and the likely effectiveness and side effects of various treatments also need to be considered \cite{Peltier_2001}.
Furthermore, identifying and managing eID risks and potential impacts require a broad set of perspectives and actors across the eID lifecycle. However, the government owns and manages the infrastructure of these systems and should be prepared to respond to risks by implementing appropriate risk-management strategies. These risk management strategies should be regularly reviewed and updated to effectively address changing threats and risks. This can involve conducting periodic risk assessments to identify new or emerging risks and reviewing and updating existing risk management strategies to ensure that they remain effective.

\noindent\textbf{Varied risk perception:}
Lay people tend to rate higher risks related to dread (e.g., catastrophes) than domain experts, who understand the evidence regarding safety limitations and controls for such systems \cite{doi:10.1126/science.3563507}. 
Moreover, people's assessment of risk is driven by their feelings and influenced by their concerns, as they naturally feel safe in their own area and are wary of danger outside of it \cite{schneier_2006}.
This creates a mismatch between the perceived and actual risks, necessitating effective risk management through structured assessment methods. As the current eID system lacks a mature risk framework and practice, having a risk assessment framework could help users systematically assess eID system risks, ensuring that the limited resources can be targeted at the highest priority risks \cite{doi:10.1080/10429247.2013.11431973}.

\noindent\textbf{Limitations:}
One key limitation of this study is that it is based on elements previously described in the literature and has considered eID systems from a generic perspective.
This may have resulted in significant issues being ignored or downplayed.
Existing literature \cite{doi:10.1089/bio.2016.0048, https://doi.org/10.1002/eet.507} recommended stakeholders' engagement to lessen the likelihood of such issues.
As part of our future work, we aim to incorporate stakeholder opinions and empirically validate the framework to provide more specific insights and ensure that broader stakeholder concerns are considered.
Lastly, risk assessments alone do not necessarily reveal appropriate mitigation strategies. In terms of future research, we hope to investigate how to identify and implement appropriate risk-mitigating strategies in eID systems.

\section{Related Work}
\label{sec:relatedwork}

\noindent \textbf{Research on Risk Assessments:} 
There are many risk assessment frameworks and standards \cite{ENISA_risk}, with most emerging from public institutions, and government bodies \cite{AlAhmad2013AddressingIS}.

\noindent The European Telecommunications Standards Institute (ETSI) \cite{ETSI} offers a threat, vulnerability, and risk assessment (TVRA) method to deal with security issues in the telecommunications industry. TVRA identifies threats to critical assets and how they can affect their operations. It also identifies the best way to mitigate these threats according to current capabilities and resource requirements.
The work in \cite{WULAN2012739} presented a framework for risk assessment within enterprise collaboration that identifies different levels of risk throughout the lifecycle of a collaborative enterprise, including pre-creation, creation, operation, and termination. The probability and impact of each risk are then determined using fuzzy linguistic terms.

The authors in \cite{10.1145/2899015.2899018} proposed a risk assessment framework for automotive embedded systems and demonstrated its viability in an industry-use case. The framework begins with a threat analysis to identify the assets and threats to those assets before estimating their threat level and impact. In another study, \cite{SAMANTRA20144010} presented a quantitative risk-assessment methodology for information technology outsourcing. The authors introduced four major steps for the method: (i) identifying the risks within the context of information technology outsourcing, (ii) collecting linguistic data about the likelihood and impact of risks from experts' opinions, (iii) multiplying the likelihood and impact of each risk, and (iv) making action plans to deal with it. 
However, despite the various existing risk assessment frameworks, the specific attributes of eID systems 
prevents the straightforward application of these risk assessment methods to eID systems. 
In addition, many of these frameworks focus primarily on general principles and guidelines, leaving users in need of more detailed implementation information \cite{5476890}.

\noindent \textbf{Research on cyber impacts:} 
Various attempts have been made to define the impact of cyberattacks on sociotechnical systems. For example, to understand how impact manifests within and outside of cyberspace, Agrafiotis et al. \cite{10.1093/cybsec/tyy006} proposed five different taxonomy of cyber harms, namely: i) physical or digital, ii) social and societal, iii) economic, iv) reputation and v) psychological. The authors also presented an initial set of metrics and methods to assess cyber harm in national contexts.
Similarly, the authors in \cite{Evangelidis2004FRAMESA} identifies five impact areas that surround e-government projects: i) economic, ii) political, iii) security, iv) societal, and v) technical risk impact areas. 
In addition, the researchers in \cite{10.1145/2899015.2899018} identified four impact areas for risk assessment in automotive embedded systems: safety, privacy and legislative, financial, and operational.
Even with the considerable research on cyber harm, no previous study has investigated the key impact areas to consider when evaluating the impact of risk on stakeholders of eID systems.

\section{Conclusion}
\label{sec:conclusions}
eID systems are crucial for achieving political, economic, and human development goals. They provide citizens with a means to prove their identity and gain access to services, thereby supporting the reduction of societal inequality. In this study, we posit that current risk assessments do not address risk factors for all key stakeholders, we explore this and analyse how potential compromise could impact them each in turn. In examination of the broader impact of risks and the potentially significant consequences for individuals, communities, and societies our framework considers a wide range of factors. Social, economic, and political contexts in which these systems were implemented are addressed and together provide a holistic platform on which to better assess the risk to eID system.

\section{Acknowledgments}
This work was supported, 
by the Bill \& Melinda Gates Foundation [INV-001309]. Under the grant conditions of the Foundation, a Creative Commons Attribution 4.0 Generic License has already been assigned to the Author Accepted Manuscript version that might arise from any submission.



\end{document}